\documentclass[notitlepage]{revtex4-1}
\usepackage{amsmath}
\usepackage{graphicx}
\usepackage{hyperref}
\usepackage[english]{babel}
\usepackage{hyphenat}
\usepackage{lipsum}

\newcommand{\M}{\mathcal{M}}
\newcommand{\GF}{G_\mathrm{F}}

\newcommand{\MeV}{\mathrm{MeV}}

\newcommand{\Mpsi}{M_{\psi}}
\newcommand{\MBc}{M_{B_c}}

 \graphicspath{{./figs/}}

\begin{document}

\author{A.~V.~Luchinsky}
\email{alexey.luchinsky@ihep.ru}
\affiliation{Institute for High Energy Physics, Protvino, Russia}
\title{Excited $\rho$ mesons in $B_{c}\to\psi^{(')}KK_{S}$ decays}
\begin{abstract}
In the presented paper exclusive decays $B_{c}\to J/\psi KK_{S}$ and $B_{c}\to\psi(2S)KK_{S}$ are analyzed. It is shown that contributions of the excited $\rho$ mesons should be taken into account to describe these decays. It is also shown that, unlike the corresponding $\tau$ lepton decays, peaks in $m_{KK_{S}}$ distributions caused by these resonances are clearly seen and can be easily separated. Theoretical predictions for the branching fractions of the reactions and $m_{\psi K}$ distributions are also presented.
\end{abstract}
\maketitle

\section{Introduction}
\label{sec:introduction}

The lightest vector hadron, i.e. $\rho(770)$ meson, has been studied in details. One cannot say the same, however, about it's excited partners, $\rho(1450)$, $\rho(1570)$, and $\rho(1700)$. For these mesons only neutral states were observed, mainly in $ee$ and $\pi\pi$ channels. Their decays into $KK$ pair is hard to detect.

One of the reactions that can be used to observe $KK$ decay of the charged excited $\rho$ meson is the $\tau$ lepton decay $\tau\to\nu_{\tau}KK_{S}$. This process was first studied experimentally by the CLEO collaboration in 1996 \cite{Coan:1996iu}. Recently a more detailed result, obtained by BaBar collaboration, appeared in \cite{BaBar:2018qry, Serednyakov:2018zck}. According to papers \cite{Bruch:2004py} CEO data can be explained theoretically using Flatte formalism \cite{Flatte:1976xu} and taking into account contributions of three $\rho$ mesons. It should be interesting to check this approach on new BaBar data.

There is, however, a fundamental problem with using $\tau$ lepton decays to analyze contributions of the excited $\rho$ mesons. It is evident that in this reaction the available energy is limited by the mass of $\tau$ lepton, $m_{\tau}=1.77$ GeV, and, for example, $\rho(1570)$ can hardly be observed. It is clear, on the other hand, that a larger energy range is available in the decays of the heavier particles, e.g. $B_{c}$ meson. In a series of papers (see, for example,
\cite{Likhoded:2009ib, Likhoded:2010jr, Berezhnoy:2011nx, Wang:2012vna, Luchinsky:2012rk, Luchinsky:2013yla}%
) it was shown how the QCD factorization theorem can be used to connect differential branching fraction of light mesons' production in exclusive $\tau$ lepton and $B_{c}$ meson decays. Predictions presented in these article are in good agreement with experimental results \cite{LHCb:2012ag, Aaij:2013gia, Aaij:2014bla, Khachatryan:2014nfa, Aaij:2016rks}. It could be interesting to try such an approach for $B_{c}\to J/\psi KK_{S}$ and $B_{c}\to\psi(2S)KK_{S}$ decays.

The rest of the paper is organized as follows. In the next section we use data on $\tau\to\nu_{\tau}KK_{}$ decay obtained by CLEO collaboration to determine the coupling constants of the excited $\rho$ mesons decays into $KK_{S}$ pair. In section \ref{sec:b_c} these results are used to make theoretical predictions for the branching fractions of $B_{c}\to\psi^{(')}KK_{S}$ decays and distributions over different kinematical variables. Short discussion is presented in the last section.

\section{$\tau\to\nu_\tau KK_{S}$ Decays}
\label{sec:tau}

\begin{figure}
  \centering
  \includegraphics[width=0.5\textwidth]{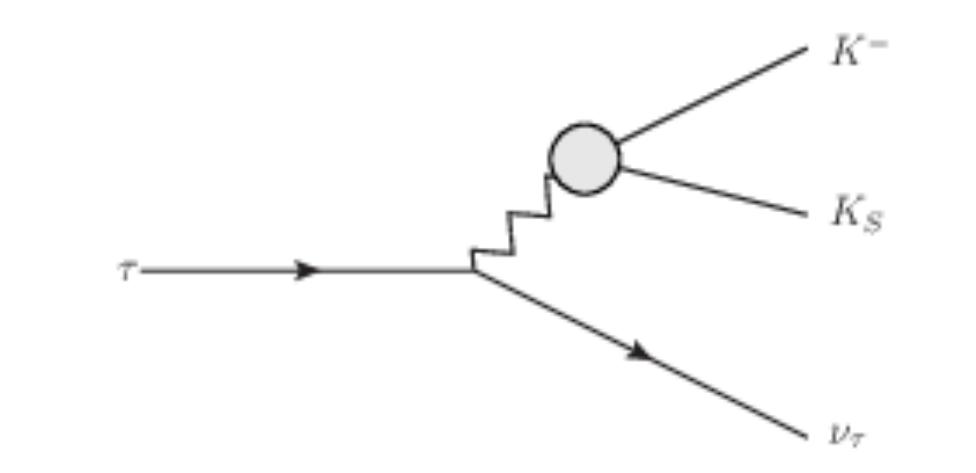}
  \caption{Feynman diagram for $\tau\to\nu_\tau KK_S$ decay}
  \label{fig:diag_tau}
\end{figure}

Let us first consider $KK_{S}$ pair production in $\tau$ lepton decay $\tau\to\nu_{\tau}KK_{S}$. The Feynman diagram describing this process is shown in Fig.~\ref{fig:diag_tau} and the corresponding amplitude can be written in the form
\begin{align}
  \label{eq:matrTau}
  \M_{\tau} &= \frac{\GF}{\sqrt{2}} \bar{u}_{\nu}(k)\gamma^{\mu}(1+\gamma_{5})u_{\tau}(P) F(q^{2})(p_{1}-p_{2})_{\mu}
\end{align}
where $P$, $k$, $p_{1,2}$ are the momenta of the initial lepton, $\tau$ neutrino and final $K$ mesons respectively (in the following we will neglect the difference in $K$ and $K_{S}$ masses), $q=p_{1}+p_{2}$ is the momentum of the virtual $W$ boson, and $F(q^{2})$ is the form factor of $W\to KK_{S}$ transition. It is clear, that the quantum numbers of final $KK_{S}$ pair should be equal to $^{G}I(J^{P})=1^{+}(1^{-})$, so this transition should be saturated by contributions of the charged $\rho$ meson and its excitations. It is convenient to use the Flatte parametrization of the form factor \cite{Flatte:1976xu} and write it in the form 
\begin{align}
  \label{eq:F}
  F(s) &= \sum_{i} c^{K}_{i} BW_{i}(s),
\end{align}
where the summation is performed over the intermediate $\rho$ mesons (in the following we will take into account only contributions of the ground state $\rho(770)$ and two excited mesons $\rho'=\rho(1450)$ and $\rho''=\rho(1700)$), $c_{i}^{K}$ are the coupling constants,
\begin{align}
  \label{eq:BW}
  BW_{i}(s) &=\frac{m_{i}^{2}}{m_{i}^{2}-s-i\sqrt{s}\Gamma_{i}(s)},
\end{align}
$m_{i}$ is the mass of the corresponding particle, and $\Gamma_{i}(\rho)$ is the energy dependent width of $\rho\to2\pi$ decay. Since final $\pi$ mesons in these decays are in $P$ wave state, the latter width can be calculated as
\begin{align}
  \label{eq:gammaRho}
  \Gamma_{i}(s) &= \frac{m_{i}^{2}}{s}\left(\frac{1-4m_{\pi}^{2}/s}{1-4m_{\pi}^{2}/m_{i}^{2}}\right)^{3/2}\Gamma_{i}
\end{align}
where $\Gamma_{i}=\Gamma_{i}(m_{i}^{2})$ is the decay widths of the corresponding meson on its mass shell.

\begin{figure}
  \centering
  \includegraphics[width=\textwidth]{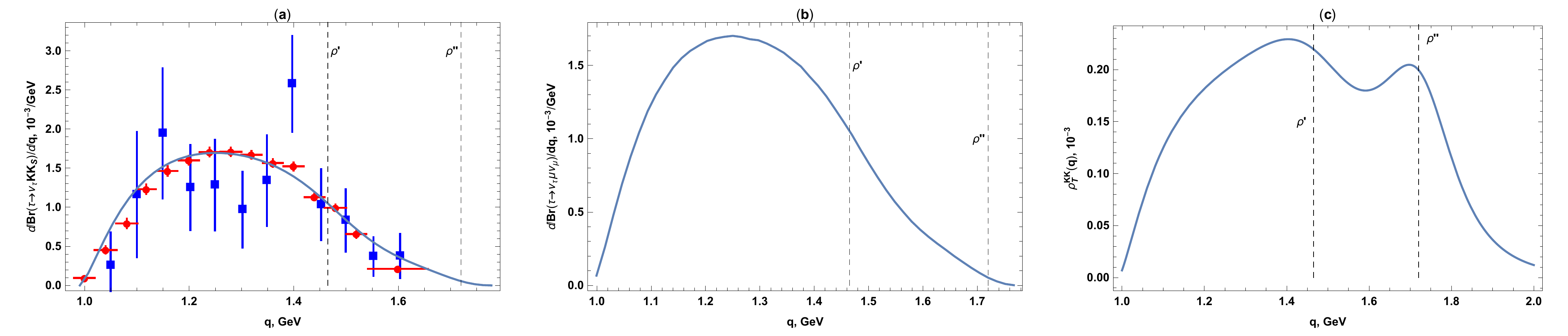}
  \caption{Differential width of $\tau\to\nu_\tau KK_{S}$ decay}
  \label{fig:dGtau}
\end{figure}

The model parameters $m_{i}$, $\Gamma_{i}$, and $c^{K}_{i}$ can be determined from analysis of the experimental data, especially $q$-distributions in the considered decay. If we are interested only in $q$ distributions of the considered decays, we can use formalism described in \cite{Tsai:1971vv}. In this framework the differential width of $\tau\to\nu_{\tau}KK_{S}$ decay can be written as
\begin{align}
  \label{eq:dGammaTau}
  \frac{d\Gamma(\tau\to\nu_{\tau}KK_{S})}{d\sqrt{q^{2}}} &= 2\sqrt{q^{2}}\frac{\GF^{2}}{16\pi m_{\tau}}\frac{(m_{\tau}^{2}-q^{2})^{2}}{m_{\tau}^{3}}(m_{\tau}^{2}+2q^{2}) \rho_{T}(q^{2}),
\end{align}
where
\begin{align}
  \label{eq:rhoT}
  \rho_{T}(q^{2}) &=\left(1-\frac{4m_{K}^{2}}{q^{2}}\right)^{3/2}\frac{\left|F(q^{2})\right|^{2}}{48\pi^{2}}
\end{align}
is the transversal spectral function of $W\to KK_{S}$ transition. 
Experimental analysis of the considered decay was performed, for example, by CLEO \cite{Coan:1996iu} and BaBar \cite{Serednyakov:2018zck,BaBar:2018qry} collaborations. In paper \cite{Bruch:2004py} obtained by CLEO collaboration results were used to determine the values of the model parameters $m_{i}$, $\Gamma_{i}$, and $c^{K}_{i}$. According to this paper in order to describe CLEO results the following values of the parameters should be used
\begin{align}
  \label{eq:params}
  m_{\rho} = 775\,\MeV, \qquad \Gamma_{\rho}=150\,\MeV,\qquad c^{K}_{\rho}=1.195\pm0.009\\
  m_{\rho'} =  1465\,\MeV, \qquad \Gamma_{\rho'}=400\,\MeV,\qquad c^{K}_{\rho'}=-0.112\pm0.010\\
  m_{\rho''} =  1720\,\MeV, \qquad \Gamma_{\rho''}=250\,\MeV,\qquad c^{K}_{\rho''}=-0.083\pm0.019.  
\end{align}
In the left panel of Fig.~\ref{fig:dGtau} we show the resulting $q$-dependence of the the differential width in comparison with obtained by CLEO and BaBar collaboration experimental results and It is clear that the agreement with this results is pretty good. The contributions of the exited $\rho$ mesons (especially $\rho''$ one), however, can hardly be seen since these mesons lie almost on the upper limit of the allowed phase space. Indeed, the relation (\ref{eq:dGammaTau}) is universal and only spectral function depends on the final hadronic state, so this relation can be rewritten in the form
\begin{align}
 \frac{d\Gamma(\tau\to\nu_{\tau}KK_{S})}{d\sqrt{q^{2}}}&=  \frac{d\Gamma(\tau\to\nu_{\tau}\mu\nu_{\mu})}{d\sqrt{q^{2}}}\frac{\rho_{T}(q^{2})}{\rho^{\mu\nu}_{T}(q^{2})},
\end{align}
where the transverse spectral function of the leptonic pair is $\rho^{\mu\nu}_{T}(q^{2})=1/(16\pi^{2})$. Transferred momentum distribution of the semileptonic $\tau$ decay is shown in figure \ref{fig:dGtau} and it is clearly seen that in the region of excited $\rho$ mesons  is strongly suppressed. That is why it could be interesting to study production of $KK_{S}$ pair in some other experiments. In the next section we will perform the calculation of $B_{c}\to\psi^{(')}KK_{S}$ decays and show that in this case the contributions of the excited states are much more clear.

\section{$B_{c}\to\psi^{(')} K K_{S}$ Decays}
\label{sec:b_c}

\begin{figure}
  \centering
  \includegraphics[width=0.5\textwidth]{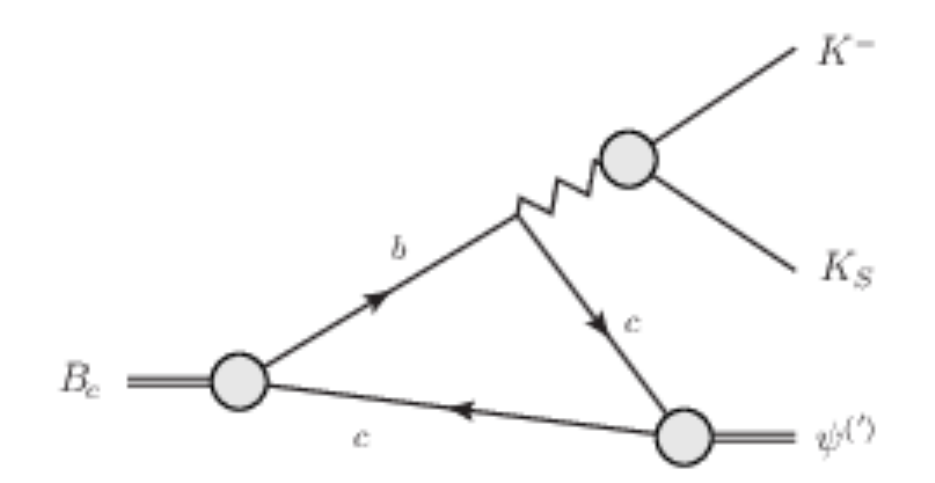}
  \caption{Feynman diagram for $B_{c}\to KK_{S}$ decay}
  \label{fig:diagBc}
\end{figure}

The decay $B_{c}\to\psi^{(')}KK_{S}$ is described by shown in Fig. \ref{fig:diagBc} Feynman diagram. The corresponding matrix element can be written as
\begin{align}
  \label{eq:matrBc}
  \M(B_{c}\to\psi^{(')}KK_{S}) &= \frac{\GF V_{cb}}{\sqrt{2}} a_{1} \langle V-A\rangle_{\mu} F(q^{2})(p_{1}-p_{2})^{\mu},
\end{align}
where $a_{1}$ is the Wilson coefficient, that describe the effect of soft gluon interaction \cite{Buchalla:1995vs}, $B_{c}\to\psi^{(')}W$ transition is described by the matrix element
\begin{align}
  \label{eq:HBC}
  \langle V-A \rangle_{\mu} &= \Big[2M_{\psi}A_{0}(q^{2})\frac{q^{\mu}q^{\nu}}{q^{2}}+
                              (M_{B_{c}}-M_{\psi})A_{1}(q^{2})\left(g^{\mu\nu}-\frac{q^{\mu}q^{\nu}}{q^{2}}\right) -
                              \nonumber\\ &
                                            A_{2}(q^{2})q^{\nu}\left(P^{\mu}+k^{\mu}-\frac{M_{B_{c}}^{2}-M_{\psi}^{2}}{q^{2}}q^{\mu}\right) -
                                            2i\frac{V(q^{2})}{M_{B_{c}}+M_{\psi}}e^{\mu\nu\alpha\beta}P_{\alpha}k_{\beta}\Big]\epsilon_{\nu},
\end{align}
where $P$, $k$, $p_{1,2}$ are the momenta of the $B_{c}$ meson, final vector charmonium, and $K$ mesons respectively, $\epsilon_{\mu}$ is the polarization vector of $\psi^{(')}$, $q=P-k$ is the momentum of virtual weak boson, $M_{B_{c}}$ and $M_{\psi}$ are the masses of the corresponding particles, and $V(q^{2})$, $A_{0,1,2}(q^{2})$ are dimensionless form factors, whose numerical values will be discussed later.

If we are interested in $q$ distribution only, we can use the formalism of the spectral functions and the corresponding decay width is equal to
\begin{align}
  \label{eq:dGBc}
  \frac{d\Gamma(B_{c}\to\psi^{(')}KK_{S})}{dq^{2}} &=
                \frac{\GF^{2}V_{cb}^{2}a_{1}^{2}\rho_{T}(q^{2})}{128\pi\MBc\Mpsi^{2}(\MBc+\Mpsi)^{2}}\sqrt{1-\frac{(\Mpsi+q)^{2}}{\MBc^{2}}}\sqrt{1-\frac{(\Mpsi-q)^{2}}{\MBc^{2}}}\times
                \nonumber\\ &
                              \Big[\Delta_1^4 \left(\Delta_1^4 A_2 ^2\left(q^2\right)+8 \Mpsi^2 q^2 V^{2}\left(q^2\right)\right)+
                              \nonumber\\ &
                              \Delta_2^4 (\MBc+\Mpsi)^4 A_1^{2}\left(q^2\right)-2 \Delta_3^6 (\MBc+\Mpsi)^2 A_1\left(q^2\right) A_2\left(q^2\right)\Big],
\end{align}
where
\begin{align}
  \label{eq:DD}
  \Delta_{1}^{4} &= \MBc^4-2 \MBc^2 \left(\Mpsi^2+q^2\right)+\left(\Mpsi^2-q^2\right)^2\\
  \Delta_{2}^{4} & = \MBc^4-2 \MBc^2 \left(\Mpsi^2+q^2\right)+\Mpsi^4+10 \Mpsi^2 q^2+q^4,\\
  \Delta_{3}^{6} &= \MBc^6-3 \MBc^4 \left(\Mpsi^2+q^2\right)+\MBc^2 \left(3 \Mpsi^4+2 \Mpsi^2 q^2+3 q^4\right)-\left(\Mpsi^2-q^2\right)^2 \left(\Mpsi^2+q^2\right)
\end{align}

\newcommand{\Br}{\mathrm{Br}}

\begin{figure}
  \centering
  \includegraphics[width=0.9\textwidth]{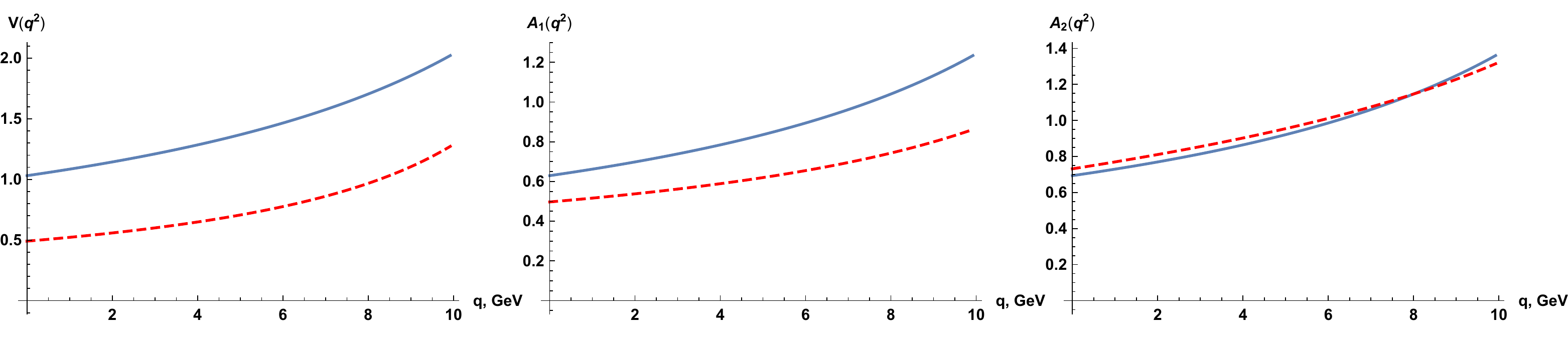}
  \caption{$B_{c}\to J/\psi W$ form factors. Solid blue and dashed red lined correspond to SR and PM form factor sets respectively}
  \label{fig:BcFF}
\end{figure}

\begin{figure}
  \centering
  \includegraphics[width=\textwidth]{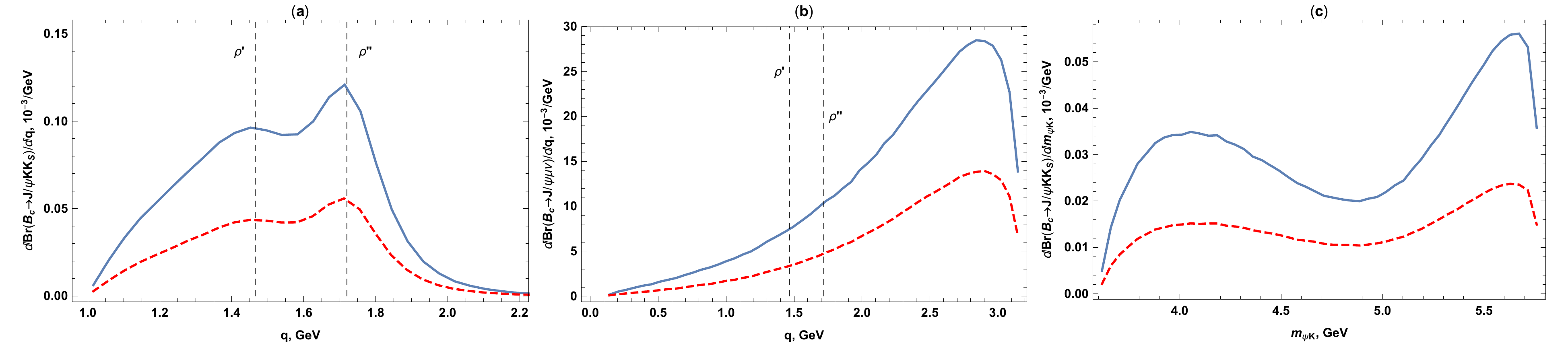}
  \caption{$B_{c}\to J/\psi\mu\nu$ and $B_c\to J/\psi KK_{S}$ distributions. Solid blue and dashed red lines correspond to SR and PM form factor sets respectively. Vertical dashed lines show the position of excited $\rho$ resonances}
  \label{fig:BcPsiKKsDistr_q}
\end{figure}

Let us discuss the parameterizations of the $B_{c}\to J/\psi KK_{S}$ decay first. It is clear that the corresponding form factors are essentially non-perturbative, so some other methods such as QCD sum rules of Potential Models should be used for their calculation. This topic is widely discussed in the literature. In the following we will use the results presented in works \cite{Kiselev:2002vz} (QCD sum rules were used in this work, in the following we will refer to it as SR) and \cite{Ebert:2003cn} (in this case the author use potential model, PM in the following). It is clear that $A_{0}(q^{2})$ form factor does not give contributions to the process under consideration. Transferred momentum dependence of all other form factors for  models used in our work is shown in figure \ref{fig:BcFF}. Using these values it is easy to see that the branching fractions of the decay in different form factors models are equal to
\begin{align}
  \label{eq:brPsiKKs}
  \Br_{SR}(B_{c}\to J/\psi KK_{S})&= (6.9\pm0.1)\times 10^{-5},\\
  \Br_{PM}(B_{c}\to J/\psi KK_{S}) &= (3.1\pm0.05)\times 10^{-5},
\end{align}
where the uncertainty is caused by the experimental error in $\tau\to\nu_{\tau}KK_{S}$ branching fractions \cite{Serednyakov:2018zck,BaBar:2018qry}. 
The corresponding $\sqrt{q^{2}}$ distributions are shown in figure \ref{fig:BcPsiKKsDistr_q}(a). One can see that, unlike $\tau\to \nu_{\tau}KK_{S}$ decay, the contributions of the excited $\rho$ mesons are clearly seen and can be easily separated. It is because in the case of $B_{c}$ meson decay the branching fraction of the semileptonic reaction $B_{c}\to J/\psi\mu\nu$ is not suppressed in $q\sim m_{\rho'}$ region [see figure \ref{fig:BcPsiKKsDistr_q}(b)]. It is also interesting to note that form of the distributions produced by different form factor sets is almost the same with the only difference in overall normalization. The reason is that, as it can be seen from the left panel of the  Figure. \ref{fig:BcFF},  in the significant for our task energy region SR and PM form factors are almost proportional to each other.

The distribution of the considered branching fraction over the invariant mass of $J/\psi K$ pair can also be observed experimentally. It is clear, that this distribution cannot be obtained using spectral function formalism, so we need to calculate the corresponding squared matrix element. As a result we have
\begin{align}
  \label{eq:d2G}
  \frac{d^{2}\Gamma(B_{c}\to\psi^{(')}KK_{S})}{dq^{2} dm_{\psi 1}^{2}} &=\frac{\GF^{2}V_{cb}^{2}a_{1}^{2}|F(q^{2})|^{2}}{2048\pi^{3}\MBc^{3}\Mpsi^{2}(\MBc+\Mpsi)^{2}}\Big\{
                                                                         \nonumber\\ &
                                                                                       -2 (\MBc+\Mpsi)^2  (m_{\psi 1}^2-m_{\psi 2}^2 )^2 A_1 A_2  (\MBc^2-\Mpsi^2-q^2 )-
                                                                         \nonumber\\ &
                                                                                       4 \Mpsi^2  V^2  (\MBc^4  (4 m_K^2-q^2 )-2 \MBc^2  (4 m_K^2-q^2 )  (\Mpsi^2+q^2 ) +
                                                                         \nonumber\\ &
                                                                                       4 m_K^2  (\Mpsi^2-q^2 )^2+ q^2  (m_{\psi 1}^4-2 m_{\psi 1}^2 m_{\psi 2}^2+m_{\psi 2}^4- (\Mpsi^2-q^2 )^2 ) )+
                                                                         \nonumber\\ &
                                                                                       (m_{\psi 1}^2-m_{\psi 2}^2 )^2  A_2^2  (\MBc^4-2 \MBc^2  (\Mpsi^2+q^2 )+ (\Mpsi^2-q^2 )^2 )+
                                                                         \nonumber\\ &
                                                                                       (\MBc+\Mpsi)^4  A_1^2  (- (16 m_K^2 \Mpsi^2- (m_{\psi 1}^2-m_{\psi 2}^2 )^2-4 \Mpsi^2 q^2 ) )                                                                                       
                                                                                       \Big\},
\end{align}
where $m_{\psi 1,2}^{2}=(k+p_{1,2})^{2}$ are the corresponding Dalitz variables (according to momentum conservation $q^{2}+m_{\psi 1}^{2}+m_{\psi 2}^{2}=\MBc^{2}+\Mpsi^{2}+2m_{K}^{2}$). The corresponding distribution is shown in Fig. \ref{fig:BcPsiKKsDistr_q}(c). It should be noted that two peaks in these distributions do not correspond to any resonances, but come from the form of $B_{c}\to\psi^{(')}\mu\nu$ matrix element.

\begin{figure}
  \centering
  \includegraphics[width=0.9\textwidth]{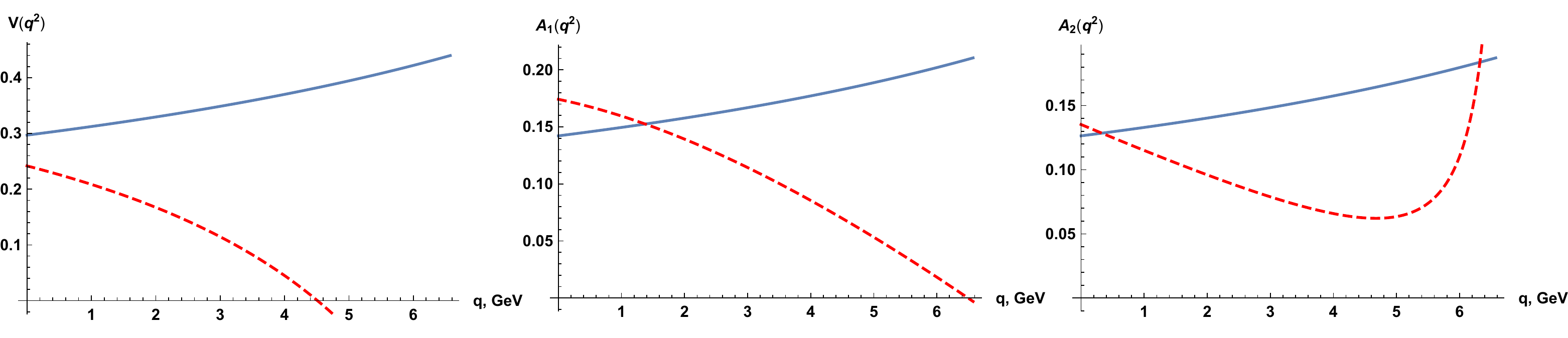}
  \caption{$B_{c}\to \psi(2S) W$ form factors. Notations are the same as in figure \ref{fig:BcFF}}
  \label{fig:BcFF2}
\end{figure}

\begin{figure}
  \centering
  \includegraphics[width=\textwidth]{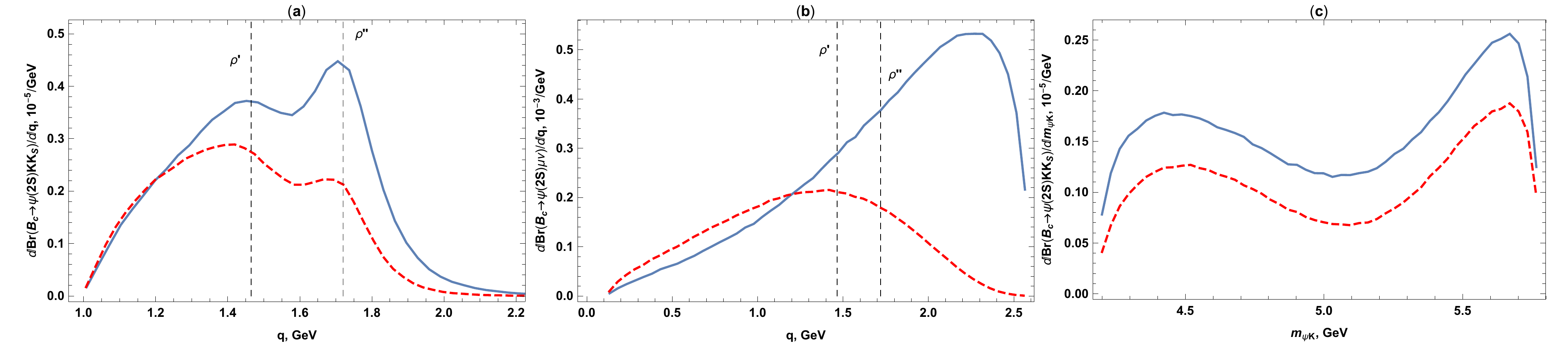}
  \caption{$B_{c}\to \psi(2S)\mu\nu$ and $B_c\to \psi(2S) KK_{S}$ distributions. Notations are the same as in Fig. \ref{fig:BcPsiKKsDistr_q}}
  \label{fig:BcPsi2SKKsDistr_q}
\end{figure}

The form factors of $B_{c}\to\psi(2S)W$ transition were also studied, for example, in papers \cite{Kiselev:2002vz,Ebert:2003cn} and we show them in figure \ref{fig:BcFF2}. Using these form factors it is easy to calculate the branching fractions of $B_{c}\to\psi(2S)KK_{S}$ decay in different models:
\begin{align}
  \label{eq:brPsi2SKKs}
  \Br_{SR}(B_{c}\to J/\psi KK_{S})&= (2.6\pm0.04)\times 10^{-6}, \\
\Br_{PM}(B_{c}\to \psi(2S) KK_{S})  &= (1.7\pm0.03)\times 10^{-6}.
\end{align}
The distributions over $KK_{S}$ and $\psi(2S)K$ invariant masses are shown in Fig. \ref{fig:BcPsi2SKKsDistr_q}. Note that in this case the forms of $q$ distributions for different form factor sets are quite different from each other.

\section{Conclusion}
\label{sec:conclusion}

In the presented article production of $KK_{S}$ pair in exclusive $\tau$ and $B_{c}$ decays is discussed. It is clear that this final state can be produced only from decay of virtual vector charged particle, i.e. $\rho$ meson and its excitations. As a result, experimental investigation of the decays can give us additional information about masses and widths of these particles and the coupling constants of $\rho^{(')}\to KK_{S}$ decays.

The decay $\tau\to KK_{S}\nu_{\tau}$ was studied experimentally, for example, in the recent BaBar papers \cite{Serednyakov:2018zck,BaBar:2018qry}. According to analysis presented in \cite{Bruch:2004py}, these results can be explained by taking into account contributions of $\rho(770)$ meson and its two excitations, $\rho(1450)$ and $\rho(1700)$. It is clear, however, that $\tau$ lepton's mass is not very large, so peak caused by the last resonance peak cannot be seen in $m_{KK_{S}}$ distribution. For this reason it could be interesting to study $KK_{S}$ pair production in decays of a heavier particle, e.g. $B_{c}$ meson.

In our paper we perform such an analysis and give theoretical description of $B_{c}\to J/\psi KK_{S}$ and $B_{c}\to\psi(2S)KK_{S}$ decays. It is clear, that the form factors of $B_{c}\to\psi^{(')}$ transitions are required for calculations of these decays, so in our work we used two different sets of these form factors, obtained using QCD sum rules and Potential models. According to our results, peaks  caused both by $\rho(1450)$ and $\rho(1700)$ resonances are clearly seen in $m_{KK_{S}}$ distributions and can be easily separated. The branching fractions of the considered decays are also calculated.

It is clear, that the final $K_{S}$ meson will be detected in $K_{S}\to\pi\pi$ decay, so observed state of the considered here decays will be $\psi^{(')}K\pi\pi$. According to \cite{Luchinsky:2013yla} the same final state can be produced also in the decay chain $B_{c}\to \psi^{(')}K_{1}\to\psi^{(')}K\rho\to\psi^{(')}K\pi\pi$ and the branching fractions of these reactions are significantly larger than the branching fractions of the decays considered in our article. It should be noted, however, that the same can also be said about the corresponding $\tau$ lepton decays, but both decays modes were observed.

The author would like to thank A.K. Likhoded and Dr. Filippova for fruitful discussions.  The work was carried out with the financial support of RFFBR (grant 19-02-00302). 


%

\end{document}